\documentclass[prl]{revtex4}
\usepackage{graphicx}
\usepackage{bm}

\begin{document}
\title{Extended RPA with ground-state correlations in a solvable model}

\author{S. Takahara, M. Tohyama}
\affiliation{Kyorin University School of Medicine, Mitaka, Tokyo
  181-8611, Japan}
\author{P. Schuck}
\affiliation{Institut de Physique Nucl$\acute{e}$aire, IN2P3-CNRS,
  Universit$\acute{e}$ Paris-Sud, F-91406 Orsay Cedex, France}

\date{\today}


\begin{abstract}
   
The ground states and excited states of the Lipkin model hamiltonian are calculated using 
a new theoretical approach which has been derived from an extended time-dependent Hartree-Fock theory 
known as the time-dependent density-matrix
theory (TDDM). TDDM enables us to calculate correlated ground states, and its small amplitude limit (STDDM),
which is a version of extended RPA theories based on a correlated ground state, can be used to calculate excited states.
It is found that this TDDM plus STDDM approach
gives much better results for both the ground states and the excited states than the Hartree-Fock ground state plus 
RPA approach.

\end{abstract}

\maketitle

The random phase approximation (RPA)
based on the Hartree-Fock (HF) ground state and the quasi-particle RPA (QRPA) based on 
the Hartree-Fock-Bogoliubov (HFB) ground state when pairing correlations are important
are standard microscopic and self-consistent theories which have extensively been used 
to study nuclear collective excitations \cite{Peter}. 
RPA and QRPA are regarded as mean-field theories because they can be formulated 
as the small amplitude limits of the
time-dependent HF theory (TDHF) and the time-dependent HFB theory, respectively.
On the other hand the time-dependent density-matrix theory (TDDM) \cite{GT90} which we have developed
in the past decade incorporates the
effects of two-body correlations into TDHF, and its small amplitude limit (STDDM) \cite{TG89}
becomes a natural extension of RPA which includes two-body amplitudes and also the effects of ground-state
correlations: The correlated ground state is obtained as a stationary solution of the TDDM equations.
We have recently applied this TDDM plus STDDM approach to study the ground states and 
low-lying quadrupole states in
unstable oxygen isotopes \cite{TS2,TTS}.  Although those results are quite encouraging, the
validity of the TDDM plus STDDM approach should be tested in further applications.
In this letter, we apply TDDM and STDDM to the Lipkin model and calculate the
ground states and excited states. We will show that TDDM and STDDM
give much better results than HF and RPA.


The correlated ground state $\vert \Phi_0\rangle$ in TDDM is
constructed such that the two equations,
\begin{equation}
\label{eqn:f1}
F_1(\alpha\beta) = \langle\Phi_0\vert [a_{\alpha}^{+}a_{\beta}, H]
\vert\Phi_0\rangle = 0,
\end{equation}
\begin{equation}
\label{eqn:f2}
F_2(\alpha\beta\alpha'\beta') = \langle\Phi_0\vert
[a_{\alpha}^{+}a_{\beta}^{+}a_{\beta'}a_{\alpha'},H]\vert \Phi_0 \rangle = 0, 
\end{equation}
are satisfied, where $[ \ \ ]$ is the commutation relation and $H$ is
the total hamiltonian consisting of the kinetic energy term and a two-body
interaction. Eqs.(\ref{eqn:f1}) and (\ref{eqn:f2}) describe the
conditions that the occupation matrix
$n^{0}_{\alpha\beta}=\langle\Phi_0\vert
a_{\beta}^{+}a_{\alpha}\vert\Phi_0\rangle$ 
and the two-body correlation matrix 
$C^0_{\alpha\beta\alpha'\beta'}=\langle\Phi_0\vert
a_{\alpha'}^{+}a_{\beta'}^{+}a_{\beta}a_{\alpha}\vert\Phi_0\rangle-{\cal
  A}(n_{\alpha\alpha'}^0 n_{\beta\beta'}^0)$, 
where ${\cal A}$ is the antisymmetrization operator, are
time-independent. We choose the single-particle wavefunction $\psi_\alpha$
as an eigenstate of the mean-field hamiltonian $h_0(\rho_0)$ which is a functional of
the one-body density matrix $\rho_0$. The expressions for 
Eqs.(\ref{eqn:f1}) and (\ref{eqn:f2}) are given in 
Ref.\cite{TTS}. We have shown in Ref.\cite{TTS} that Eqs.(\ref{eqn:f1}) and (\ref{eqn:f2}) can be solved
using the iterative gradient method: Starting from the 
HF ground state where $n^0_{\beta\alpha}=\delta_{\beta\alpha} (0)$
for an occupied (unoccupied) state and
$C^0_{\alpha\beta\alpha'\beta'}=0$, we iterate
\[
\left(
\begin{array}{c}
n^0(N+1)\\
C^0(N+1)
\end{array}
\right)
=
\left(
\begin{array}{c}
n^0(N)\\
C^0(N)
\end{array}
\right)
-\alpha\left(
\begin{array}{cc}
a & c\\
b & d
\end{array}
\right)^{-1}
\left(
\begin{array}{c}
F_{1}(N)\\
F_{2}(N)
\end{array}
\right)
\]
until convergence is achieved. The matrix elements $a, b, c,$ and $d$,
which are functional derivatives of $F_1$ and $F_2$ and depend on the
iteration step $N$, are equivalent to those
appearing in the hamiltonian matrix of STDDM. They are also given in
Ref.\cite{TTS}. We have to introduce a
small parameter $\alpha$ to control the convergence process.


Excited states are obtained solving the STDDM equations for the
one-body transition amplitude 
$x_{\alpha\beta}(\mu)=\langle\Phi_0\vert a^{+}_\beta 
a_\alpha\vert\Phi_{\mu}\rangle$
and the two-body transition amplitude 
$X_{\alpha\beta\alpha'\beta'}(\mu)=\langle\Phi_0\vert
a^{+}_{\alpha'}a^{+}_{\beta'} a_{\beta} a_{\alpha}\vert\Phi_{\mu}\rangle$, 
where $\vert\Phi_{\mu}\rangle$ is the wavefunction for an excited state with
excitation energy $\omega_\mu$. The equations in STDDM can be written in matrix form
\begin{eqnarray}
\left(
\begin{array}{cc}
a & c \\
b & d
\end{array}
\right)\left(
\begin{array}{c}
x \\
X
\end{array}
\right)
=\omega_\mu
\left(
\begin{array}{c}
x \\
X
\end{array}
\right).
\label{eq:STDDM}
\end{eqnarray}
When the ground-state $\vert\Phi_0\rangle$ is approximated by the HF one
and only the particle (p) - hole (h) and 2p - 2h amplitudes (and their
complex conjugates) are taken in Eq.(\ref{eq:STDDM}), STDDM reduces
to the second RPA (SRPA) \cite{SRPA1}. 
The strength function $S(E)$ defined as
\begin{equation}
S(E)=\sum_{\omega_\mu > 0}\vert \langle \Phi_\mu \vert 
\hat{Q}\vert\Phi_0\rangle\vert^2\delta(E-\omega_\mu)
\end{equation}
for an excitation operator $\hat{Q}$ is given in terms of the solution
of Eq.(\ref{eq:STDDM}).
The detailed expressions of $S(E)$ for one-body and two-body excitation operators are given in Ref.\cite{TTS}.

The Lipkin model \cite{Lip} describes an N-fermions system with two
N-fold degenerate levels with energies $\epsilon/2$ and $-\epsilon/2$,
respectively. The upper and lower levels are labeled by quantum number
$p$ and $-p$, respectively, with $p=1,2,...,N$.  The model is
described by the following hamiltonian
\begin{equation}
H=\epsilon J_{z}+\frac{V}{2}(J_+^2+J_-^2),
\end{equation}
where the operators are given as
\begin{eqnarray}
J_z=\frac{1}{2}\sum_{p=1}^N(a_p^{+}a_p-{a_{-p}}^{+}a_{-p}) \\
J_{+}=J_{-}^{+}=\sum_{p=1}^N a_p^{+}a_{-p}.
\end{eqnarray}
Using an $N=4$ system as an example, we solve Eqs.(\ref{eqn:f1}) and (\ref{eqn:f2}) for the
ground states and Eq.(\ref{eq:STDDM}) for the excited states.
The hamiltonian matrix in STDDM (Eq.(\ref{eq:STDDM})) is not hermitian in general \cite{TTS,TS1} and, as a
result, there is a possibility that some eigenvalues become complex. However, the hamiltonian in the
Lipkin model is so simple that all eigenvalues are real and that the strength
functions are positive definite.

\begin{figure}
\begin{center}
\includegraphics[height=6cm]{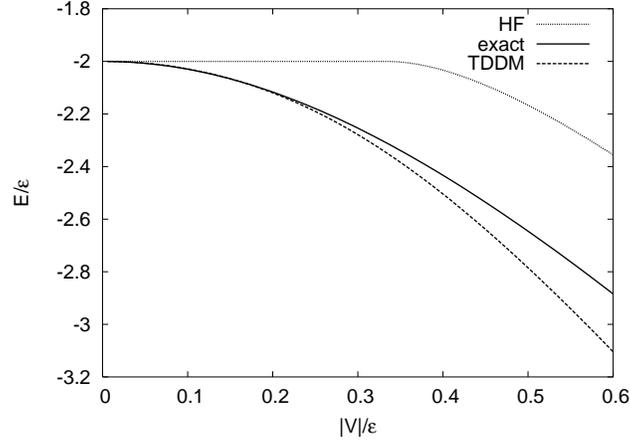}
\end{center}
\caption{Total energy $E/\epsilon$ as a function of $\vert V \vert$
  for $N=4$.  The solid, dashed, and dotted lines depict the exact
  solutions, the TDDM results, and the HF results, respectively.}
\label{fig:gs}
\end{figure}

First we present the results for the ground states. 
We fix $\epsilon$ at 2 and change $V$.
To obtain converged results in the iterative gradient method, we need 
to start with a small value of interaction strength $V$ and gradually
increase it: We start with $V/50$ and increase it by
$V/50$ for each 200 iterations. We found that 
the correlated ground states obtained coincide with those
calculated using a time-dependent approach \cite{To94} which provides us with another method for obtaining a stationary
solution of the TDDM equations.
The total energies obtained in TDDM (dashed line) for various interaction strength
$\vert V \vert$ are shown in Fig. \ref{fig:gs} in comparison with the
exact (solid line) and HF (dotted line) ones.  The Lipkin model doesn't have stable HF ground
states for $\vert V\vert / \epsilon > 1 / (N-1)$. For $\vert V\vert/\epsilon
>1/3$ in the case of $N=4$, the 'deformed' HF solution is shown
\cite{Peter,Lip}. The results in TDDM and STDDM are always calculated using the original (not deformed) single-particle basis. 
The results in TDDM are very close to the exact ones
in the wide range of the interaction strength:
The deviation of the total energy from the exact
one is only 1.1\% at $\vert V\vert/\epsilon=0.3$.

\begin{figure}
\begin{center}
\includegraphics[height=6cm]{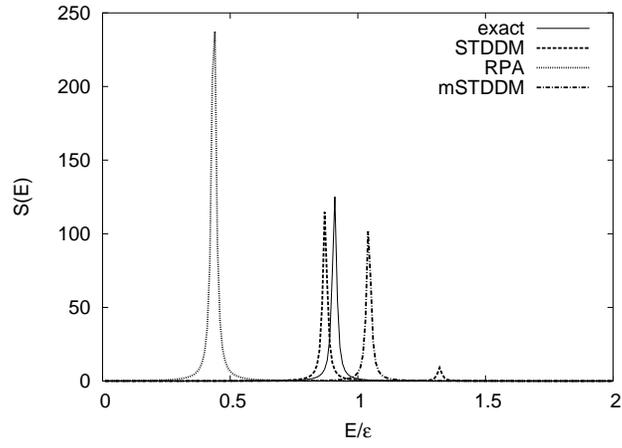}
\end{center}
\caption{Strength distributions of the one-phonon state calculated in
STDDM (dashed line) and RPA (dotted line). The exact solution is shown with the solid line.
The strength functions are smoothed with an artificial width $\Gamma/\epsilon=0.025$.
}
\label{fig:q1}
\end{figure}

\begin{figure}
\begin{center}
\includegraphics[height=6cm]{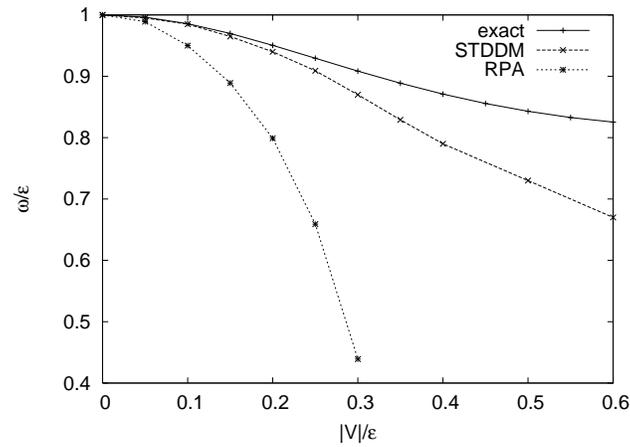}
\end{center}
\caption{Excitation energy of the first excited state as a function of the interaction strength.
The dashed and dotted lines connect the results in TDDM and RPA, respectively, while the solid line indicates the
exact solutions.}
\label{fig:q1b}
\end{figure}

Now we show the results for the first excited state (one-phonon state). 
We use $V=-0.6$ which is slightly smaller than the critical value
$V=-2/3$.
Fig. \ref{fig:q1} shows the strength function for
the excitation operator $\hat{Q}_1=J_{+}+J_{-}$ calculated in STDDM
(dashed line).
The strength functions for the exact solution (solid line) and the RPA solution
(dotted line) are also drawn for comparison. The dot-dashed line depicts the
result in a modified STDDM (mSTDDM) which will be discussed in connection with the
two-phonon state.
To make differences among various calculations easy to see, we smoothed
the strength functions with an artificial width $\Gamma/\epsilon=0.025$.
RPA overestimates the collectivity of the first excited state: The
excitation energy is much lower than the exact solution and the
transition strength becomes quite large.
On the other hand both the excitation energy and the transition
strength calculated in STDDM are very close to the exact values.
There appears a state at $E/\epsilon=1.3$ in the STDDM result.
This is due to the coupling to 3p-1h and 1p-3h configurations.
The value of the energy-weighted sum $m_1$ of the transition strength is 14.2 in STDDM while the exact value
is 14.6. The value of $m_1$ in RPA is 15.2.
The results in STDDM demonstrate the importance of ground-state correlations 
and the coupling to the two-body amplitudes which are completely missing in RPA. The effects of
ground-state correlations may be classified as the self-energy
contributions, the modification of p-h interactions and the vertex
corrections \cite{TSA88}. For example, the fact that the excitation energy in STDDM is much larger than
that in RPA is due to the increase in the self-energy of 1p-1h configurations. However,
it is not easy in our approach to clearly distinguish the various contributions of the ground-state correlations.
The excitation energy of the one phonon state is shown in Fig.\ref{fig:q1b} as a function of the interaction strength.
It is clearly seen that STDDM avoids the breakdown of RPA in the region of the phase transition ($\vert V\vert/\epsilon
>1/3$) by
including the effect of ground-state correlations.

\begin{figure}
\begin{center}
\includegraphics[height=6cm]{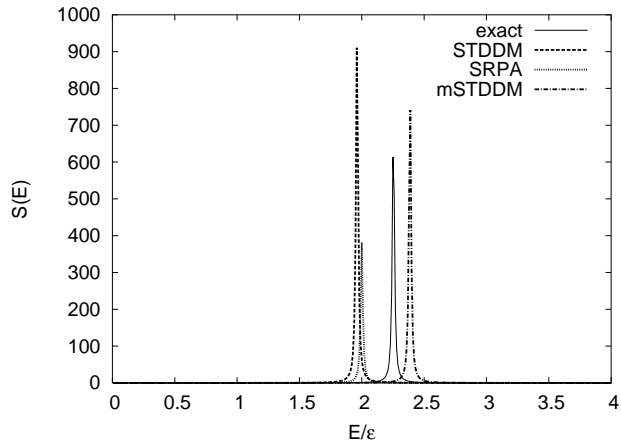}
\end{center}

\caption{Strength distribution of the two-phonon state of the Lipkin
  model calculated in STDDM (dashed line). The solid and dotted lines
  depict the exact solution and the result in SRPA, respectively. The
  result in mSTDDM is shown with the dot-dashed line.
The strength functions are smoothed with an artificial width $\Gamma/\epsilon=0.025$.
}
\label{fig:q2}
\end{figure}

The strength function in STDDM (dashed line) 
for the two-phonon state excited with
$\hat{Q}_2=(J_{+}+J_{-})^2$ is shown in Fig. \ref{fig:q2}, where the strength
function of the exact solution (solid line) and that in SRPA (dotted line) are also
drawn for comparison. The interaction strength used is $V=-0.6$.
In SRPA where only 2p-2h and 2h-2p amplitudes are considered, neither coupling
to the one-body amplitudes nor interaction among 2p-2h configurations exist
in  the Lipkin model. Therefore, the result in
SRPA is equivalent to the unperturbed one. The state in STDDM has much larger
transition strength than SRPA. This is due to the fact that
$X_{pp'p''p'''}, X_{php'h'}$ and $X_{hh'h''h'''}$ amplitudes in
addition to $X_{pp'hh'}$ and $X_{hh'pp'}$ are taken into account in
STDDM. However, the transition strength is overestimated and the
excitation energy is lower than the exact solution. This is similar to
the relation between the RPA result and the exact one for the
one-phonon state. This means that the effects of ground-state correlations are not
fully taken into account in STDDM for the two-phonon state. In fact,
the self-energy terms for 2p-2h configurations are not
included in STDDM because they are kinds of three-body effects.
In Ref.\cite{TTS} we presented a prescription for taking into account the
three-body effects in which missing terms are added to the hamiltonian matrix 
acting on the two-body amplitudes. (See Eq.(25) of Ref.\cite{TTS} for a detailed explanation of the prescription.) We call this version of 
STDDM a modified STDDM (mSTDDM). The result in mSTDDM is shown in
Fig. \ref{fig:q2} with the dot-dashed line. It becomes closer to
the exact solution: The transition strength is
reduced and the excitation energy is increased. Due to the coupling to the two phonon state, the first excited state is slightly modified as well, as
shown in Fig.\ref{fig:q1}.

In summary, we applied the TDDM plus STDDM approach to the Lipkin model. 
It was found that this approach gives much better results for the ground-state energies and
the first excited states (the one-phonon state) than HF and RPA. 
However, the second excited state (two-phonon
state) calculated in STDDM was not so good as the one-phonon state. This
is due to the lack of some ground-state correlations coming
from 3-body effects. It was shown that the modified STDDM which includes
the three-body effects can improve the result for the two-phonon state.



\end{document}